\documentclass[11pt]{article}
\usepackage{indentfirst}
\usepackage{graphicx}
\usepackage{amsmath,amsthm,amssymb,verbatim,url}
\usepackage{xifthen}
\usepackage{wasysym}
\usepackage{hyperref}
\usepackage{makeidx}
\usepackage{xifthen} 
\hypersetup{colorlinks=true,urlcolor=blue}

\newcommand{\bv}{\begin{verse}}
\newcommand{\ev}{\end{verse}}
\newcommand{\be}{\begin{equation}}
\newcommand{\ee}{\end{equation}}
\newcommand{\bea}{\begin{eqnarray}}
\newcommand{\eea}{\end{eqnarray}}
\newcommand{\bq}{\begin{quotation}}
\newcommand{\eq}{\end{quotation}}
\newcommand{\tr}{{\rm tr}}

\newcommand{\myurl}[2][]{\ifthenelse{\isempty{#1}}{\url{#2}}{\href{#1}{\tt #2}}}

\begin{document}

\begin{center}
\LARGE {\bf QBism, Polishing Some Points}\bigskip\bigskip\\
\Large Christopher A. Fuchs \bigskip \\
\Large Blake C. Stacey \bigskip \\
\normalsize Department of Physics, University of Massachusetts Boston \\ 100 Morrissey Boulevard, Boston MA 02125, USA \smallskip \\

\end{center}

\bigskip\bigskip

\bq
\noindent \small
{\bf Abstract:}
QBism\index{QBism} pursues the real by first eliminating the elements of quantum theory too fragile to be ontologies on their own. Thereafter, it seeks an ``ontological lesson'' from whatever remains. Here, we explore this program by highlighting three tenets of QBism. First, the Born Rule\index{Born Rule} is a normative statement.  It is about the decision-making behavior {\it any\/} individual agent should strive for, not a descriptive ``law of nature.''  Second, all probabilities, including all quantum probabilities, are so subjective they never tell nature what to do.  This includes probability-1 assignments.  Quantum states thus have no ``ontic hold'' on the world, which implies a more radical kind of indeterminism in quantum theory than other interpretations understand.  Third, quantum measurement outcomes {\it just are\/} personal experiences for the agent gambling upon them.  Thus all quantum measurement outcomes are local in the sense of the agent enacting them. Through these tenets, we explain four points better than previously: 1) how QBism contrasts with Bohr's \index{Bohr, Niels} concern over unambiguous language, 2) how QBism contrasts with the Everett \index{Everett, Hugh III} interpretation, 3) how QBism understands the meaning of Bell inequality\index{Bell inequality} violations, and 4) how QBism responds to Wigner's\index{Wigner, Eugene P.} ``suspended animation''\index{Wigner's friend} argument. Finally, we consider the ontological lesson of the tenets and ask what it might mean for the next one hundred years of quantum theory and humankind more generally.
\eq


\section{What is QBism?}
\label{Tenets}

QBism, \index{QBism} initially an abbreviation for Quantum Bayesianism, is now a stand-alone term for a point of view on quantum theory that goes well beyond its Bayesian roots~\cite{Fuchs10a,Fuchs14,Fuchs16,Mermin14f,Fuchs13a,Varenna17,Mermin19,Healey16,vonBaeyer16}. Applying the term ``Bayesian'' to the view grew awkward, as there are many varieties of ``objective Bayesianism'' (and even some mildly subjective ones)~\cite{Good83} to which QBism does not subscribe~\cite{Stacey16,Stacey2022,Stacey2023b}.  For a while David Mermin \index{Mermin, N. David} toyed with taking the B to stand for the B in \index{de Finetti, Bruno} Bruno de Finetti---the founder of the variety of personalist Bayesianism that QBism {\it does\/} subscribe to---but that was tongue-in-cheek~\cite{Mermin13}.  A more serious option for the B was the rather unattractive word ``bettabilitarianism'' \index{bettabilitarianism} introduced by \index{Holmes, Oliver Wendell Jr.} Oliver Wendell Holmes Jr.\ to describe his view of the universe. That one actually hits the nail on the head for QBism.  Holmes wrote~\cite[pp.~251--253, his italics, our underlining]{Holmes29}:
\bq\small
\noindent Chauncey Wright \index{Wright, Chauncey} a nearly forgotten philosopher of real merit, taught me when young that I must not say {\it necessary\/} about the universe, that we don't know whether anything is necessary or not.  So I describe myself as a {\it bet\/}tabilitarian.  I believe that we can {\it bet\/} on the behavior of the universe in its \underline{contact} with us.
\eq
``In its contact with us''---that is perhaps QBism's\index{QBism} most radical move from the common notion of physics as a description of ``what is out there without us.''   Instead, for QBism, a gambling agent who adopts quantum theory becomes a ``better bettabilitarian''~\cite{Varenna17}.  He is better able to thrive in the world in which he is immersed---a world whose characteristics cause the quantum formalism to become an advisable addition to his probability calculus.

Three characteristics set QBism\index{QBism} apart from other existing interpretations of quantum mechanics. The first is its crucial reliance on the mathematical tools of quantum information theory to reshape the look and feel of quantum theory's formal structure. The second is its stance that two levels of radical ``personalism'' are required to break the interpretational conundrums plaguing the theory. The third is its recognition that with the solution of the theory's conundrums, quantum theory does not reach an end, but is the start of an adventure: QBism is a continuing project.

The two levels of personalism refer to how the ``probabilities'' and ``measurement events'' of quantum theory are to be interpreted. With regard to quantum probabilities, QBism\index{QBism} asserts that they are to be interpreted as genuinely personal, Bayesian degrees of belief. This is the idea that probability is not something out in the world that can be right or wrong, but a personal accounting of what one expects. The implications of this are deep, for one can see with the help of quantum information theory that it means that quantum states, too, are not things out in the world. Quantum states rather represent personal accounting, and two agents speaking of the same quantum system may have distinct state assignments for it. In fact, there are potentially as many quantum states for a system as there are agents interested in considering it.

The second level of personalism appears with the meaning of a quantum-measurement outcome. On this, QBism\index{QBism} holds with Pauli (and against Bohr) that a measurement apparatus must be understood as an extension of the agent himself, not something foreign and separate. A quantum measurement device is like a prosthetic hand, and the outcome of a measurement is an unpredictable, undetermined ``experience'' shared between the agent and the external system. The latter part of this we see as the ultimate message of the Bell \index{Bell inequality} and Kochen--Specker theorems \index{Kochen--Specker theorem} \index{Kochen, Simon} \index{Specker, Ernst T.} and their variants~\cite{Cabello16,Ben-Menahem97}---that when an agent reaches out and touches the world, a little moment of creation occurs in response. Quantum theory, thus, is no mirror image of the world, for ``there is no one way the world is;'' it is ``still in creation, still being hammered out''~\cite[p.~742]{FuchsOx}.  Rather the theory should be seen as a ``user's manual'' that any agent can adopt for better coping with the world external to himself. The agent uses the manual to help guide his little part and participation in the world's ongoing creation.

When contemplating a quantum measurement, one makes a conceptual split in the world: One part is treated as an agent, and the other as a kind of reagent or catalyst (something that brings about a change in the agent). In older terms, the former is an observer and the latter a quantum system modeled by a \index{Hilbert space} Hilbert space ${\cal H}_d$ of some finite dimension $d$.\footnote{\label{FootNoteX} The restriction to finite-dimensional quantum-state spaces is not essential, but it does make the mathematics cleaner.  At times we have wondered whether this is a hint that quantum mechanics should more properly be restricted to finite spaces~\cite{Fuchs10a,Varenna17}, but this is not the place to discuss it.}  A quantum measurement consists first in the agent taking an {\it action\/} on the quantum system. The action is formally captured by a set of measurement operators $\{\hat A_i\}$ (a positive-operator valued measure; see Section \ref{QuantumInfo}). The action leads generally to an incompletely predictable consequence, a particular personal {\it experience\/} $A_k$ drawn from the set. The action is identified with the set, the experience with a single element within the set.  Furthermore, as anything in the agent's external world may be considered a quantum system, an action can be anything from running across L'Etoile de Paris (and gambling upon one's life) to a sophisticated quantum optics experiment (perhaps gambling on the violation of a \index{Bell inequality} Bell inequality).

The quantum state $|\psi\rangle$ for the system lives inside the agent's head: It captures his degrees of belief concerning the consequences of his actions upon the system, and in contrast to the system itself, has no existence in the external world. If the agent were to go poof, the quantum state would go poof too.  Quantum states are not part of the scaffolding of the world.

Quantum measurement is not a passive observation of what is out there, in the motif of an eye looking upon the world, but an action upon the world that aids it in releasing its potential. The measuring devices should be considered an integral part of the agent himself. The consequence of each quantum measurement is a unique creation within the previously existing universe.  Wolfgang Pauli \index{Pauli, Wolfgang} characterized this picture as a ``wider form of the reality concept''~\cite{Fuchs18} than that of Einstein's, \index{Einstein, Albert} which he labeled the ``ideal of the detached observer.''

Let us now elaborate on these ideas by considering three of the eight fundamental tenets expressed in Ref.\ \cite{Fuchs2023}.  In this treatment, however, we will better polish four points than previously:  1) how QBism\index{QBism} contrasts with Bohr's \index{Bohr, Niels} concern over unambiguous language, 2) how QBism contrasts with the Everett \index{Everett, Hugh III} interpretation, 3) how QBism understands the meaning of Bell inequality \index{Bell inequality} violations, and 4) how QBism responds to Wigner's ``suspended animation'' \index{Wigner, Eugene P.} argument. Finally, we describe how a QBist ontology must be more radical than \index{relationalism} \index{perspectivalism} mere ``relationalism'' or ``perspectivalism.''

\subsection{First Tenet: Quantum Theory Is Normative, Not Descriptive}
\label{FirstTenet}

If for QBism,\index{QBism} quantum theory is not a direct {\it description\/} of reality as it is, then what does it mean for a physicist or an experimenter to ``accept quantum theory''? It means to accept the formal structure of quantum theory, along with the Born Rule, as a normative aim in making decisions.  But one must be careful with regard to the meaning of the Born Rule: \index{Born Rule} In a common reading (even in early Quantum Bayesianism~\cite{Fuchs00} before it was QBism), one might say, ``To use the Born Rule means to take the quantum state implied by the preparation procedure and the projections associated with the measurement device and put them together to calculate probabilities for the measurement outcomes.''  However, in a statement like that, the state and the projections are implicitly taken to be logically prior to the probabilities calculated from them. The textbooks treat them as something more fundamental than the probabilities they give rise to, and one naturally slips into the trap of accepting that. ``You're not going to tell me that the ground state of intergalactic hydrogen is anyone's degree of belief,''\ someone once said~\cite{Ballentine03}.  Well, in QBism it is exactly that!  A quantum state is a specific agent's catalog of beliefs for the possible experiences that would arise should agent take an action on a system.  These beliefs may be widely shared by a number of agents who consider taking actions on the same system, but conceptually beliefs they remain.  What is {\it objective\/} in QBism is the standard by which all 
the beliefs should relate to each other.

For example, take a spin-1/2 particle~\cite{Caves07}. Suppose an agent contemplates acting on it in such a way that a $\hat S_x$ measurement is made.  He asks himself how he would gamble on those potential outcomes---i.e., what his expectation value $\langle S_x\rangle$ might be.  He next contemplates the same of a $\hat S_y$ measurement, and then of a $\hat S_z$ measurement. Absent the strictures of quantum theory, the agent's gambles on three different, mutually exclusive experiments need have no relation at all.

But then suppose the agent contemplates how he would gamble upon the outcomes of a spin measurement $\hat S_n$ in some fourth direction.  It is here that the Born Rule gives some guidance---for the Born Rule is an addition to the raw probability calculus. Using it signifies that the context of the probability assignments is not some general setting, but that of various hypothetical quantum measurements the agent could perform on the given system.  The three first probabilities, being {\it informationally complete}, \index{informational completeness} specify a unique density operator $\hat\rho$. This density operator then, once again through the Born Rule, will lead uniquely to the fourth set of probabilities, and consequently a fourth expectation value $\langle S_n\rangle={\rm tr}\,\hat\rho\hat S_n\,$.  That is, once the three expectations are set, the fourth has no freedom at all.  In this sense, the expectations come first and last; the density operator is only a construct in between.

However, this is all a mental game for the agent.  Probabilities are not given to him on a plate.  Ultimately, they are guesses he makes after trying to take as many considerations into account as he has time or mental capacity for.  If after calculating the fourth set of probabilities, he finds them so absurd, so discordant with how he thinks he should gamble on the outcomes of the fourth measurement, he might go back and completely readjust his expectations for $\hat S_x$ or $\hat S_y$ or $\hat S_z$ or any combination of the three.  For instance, he might do this based on what he {\it more genuinely believes\/} of a measurement of $\hat S_n$ (perhaps he has more experience making measurements in that odd direction because his lab table is a little crooked).  Nothing is sacred except that he should {\it strive\/} to satisfy the Born Rule \index{Born Rule} for all four probabilities.  Which is to say, the agent should {\it strive\/} to find a single state assignment $\hat\rho$ that encapsulates all the relevant probabilities.  

The fact that one needs \emph{three} expectation values to nail down a unique solution is a \emph{structural fact} about quantum theory, more basic than any particular choice of expectation values. Likewise for the constraint that they must lie within the unit ball: $\langle S_x\rangle^2 + \langle S_y\rangle^2 + \langle S_z\rangle^2 \leq 1$. Another theory attuned to a world of different character might demand a different number of coordinates, bounded by a different shape~\cite{Stacey2025}. We will return to this theme shortly, developing it in a clarifying way by reducing the number of measurements needed, from three to one.

But let us emphasize that, as the probabilities are adjusted and readjusted and guessed at and written hastily because the time is up for a decision, etc., the derived quantum state should not be imagined to have any objective character.  It is not that quantum states are full of human error, but that they are full of humanity.  Quantum states are mental constructs whose building blocks are themselves subjective probability assignments, expectations, and gambling commitments. Furthermore, we rely on a key distinction between the personalist notion of probability employed by QBism\index{QBism} and the objective Bayesian notion espoused by \index{Jaynes, Edwin T.} E. T. Jaynes~\cite{Jaynes03}.  Markus Mueller \index{Mueller, Markus P.} once put the issue like this~\cite{Mueller16}: 
\bq\small
\noindent
I don't buy the difference between ``objective Bayesianism'' and (de Finetti-type) \index{de Finetti, Bruno} ``personalist Bayesianism'' in general; the distinction seems artificial to me.  According to the latter viewpoint, different agents can assign different probabilities because they have different backgrounds and experiences.  Fine.  But the thing is: For {\it fixed given\/} background and experiences (including one's knowledge of one's name, one's memories, previous probability assignments, really everything), the assignment should be objective.  That is, there should be a normative assignment of probabilities $P(y|x)$ where $y$ encodes an agent's future experiences, and $x$ encodes the totality of {\it all\/} the agent's past experiences, everything she knows and sees and remembers right now. This is in line with {\it both\/} views on Bayesianism, as far as I can see. It just follows from the fact that ``you only have what you have'' (which is $x$), and you {\it must\/} assign some probability.  
\eq
But in fact, this is exactly where QBism\index{QBism} differs on things:  An agent {\it never\/} has $x$ as defined here.  Even with respect to the single agent, the probability assignment of choice is not unique.  As \index{van Fraassen, Bas C.} van Fraassen argued~\cite{Fraassen84}, a probability assignment always involves a ``leap of faith'' in the sense of William James's \index{James, William} \index{will to believe} ``will-to-believe''~\cite{James1879,James1896}.  For the mesh of beliefs simply can never be made complete and exhaustive.  If the world itself is not ``complete and exhaustive''---as the kinds of ontology QBism toys with indicate---why should anyone's mesh of beliefs be?  QBism thus interprets {\it probability theory as a whole\/} in the normative sense, but not so for any of the provisionary probability assignments within it.

\subsubsection{Generalized Measurements, Reference Measurements}
\label{QuantumInfo}
This way of viewing the Born Rule \index{Born Rule} can be made clearer and more general with the introduction of a little quantum information theory.  It is worthwhile going through this exercise to show just how distinct QBism's\index{QBism} way of viewing the Born Rule is.  At the exercise's end, the quantum state as a complex vector in Hilbert space \index{Hilbert space} and a measurement specification as a set of Hermitian operators will have completely disappeared from the formalism. Only probabilities and conditional probabilities will remain, and the normative character of the Born Rule will be seen as a particular relation an agent should strive to maintain between his probability assignments.

With the advent of quantum information theory it was realized that the von Neumann notion of a quantum measurement was too limiting for many natural problems.  Thus arose the notion of the positive-operator valued measure (or POVM) \index{positive-operator valued measure (POVM)} in the 1970s as the most general kind of quantum measurement~\cite{Nielsen10}.  In this theory, the outcomes or ``clicks'' $A_j$ of a measuring device are modeled by a {\it set\/} of positive semi-definite operators $\{\hat{A}_j\}$ on the Hilbert space ${\cal H}_d$ 
associated with the quantum system.\footnote{The $A_j$ with no hat denotes a lived experience, the click; the associated operator $\hat{A}_j$ is a tool for calculations.}
This means that the operators $\hat{A}_j$ are Hermitian with nonnegative eigenvalues and, together, constitute a resolution of the identity operator $\hat{I}$ on ${\cal H}_d$, i.e.,
\be
\sum_j \hat{A}_j = \hat{I}\;.
\ee
Beyond these conditions, however, there are no further restrictions on the operators $\hat{A}_j$.  For instance, even though the operators live on a $d$-dimensional Hilbert space, the index $j$ may take values from any range whatsoever, even a continuously infinite set.  Yet, in principle, any such set can be made to correspond to the clicks of an appropriate measuring device.

It is worth appreciating how different this notion of quantum measurement is from the notion described in the founders' papers on quantum theory.  There, a measuring device was associated with a {\it single\/} Hermitian operator and the outcomes of the measurement with the {\it eigenvalues\/} of that operator.  Consequently, for a finite dimensional space ${\cal H}_d$, a quantum measurement could have at most $d$ distinct outcomes.  In the generalized notion of measurement, the outcomes are associated not with the eigenvalues of a single operator, but each with an operator itself.  If one wanted to think of the old notion of measurement in the new terms, then the outcomes $A_j$ of a measurement would be associated with the eigenprojectors $\hat{A}_j=|j\rangle\langle j|$ of the encompassing Hermitian operator, instead of its eigenvalues.

The most important thing this buys for our purposes is that one can get a notion of {\it informational completeness\/} in the generalized setting, \index{informational completeness} just as in the previous subsection, even with a {\it single\/} quantum measurement.  Informational completeness does not necessarily rely on the consideration of some number of complementary observables, as $\hat S_x\,$, $\hat S_y\,$, $\hat S_z$ were.  In other words, one can now have a {\it single kind of quantum measurement\/} through which the outcome probabilities uniquely specify a $\hat\rho$.

The starting point is to look at how the Born Rule \index{Born Rule} gets generalized for these generalized measurements.  The answer is, it doesn't.  For a quantum state $\hat\rho$, pure or mixed, and a measuring device described by a set of operators $\{\hat{A}_j\}$, the probabilities for the individual clicks $A_j$ are given by
\be
P(A_j)= {\rm tr}\, \hat\rho\hat{A}_j \;.
\label{BR}
\ee
By the properties of the $\hat{A}_j$, one will have automatically that $P(A_j)\ge0$ and $\sum_j P(A_j) = 1$, exactly as required of a probability distribution, no matter what the density operator $\hat\rho$.  Furthermore, in the special case that the measurement is one of the old von Neumann variety, Eq.\ (\ref{BR}) reduces to $P(A_j)=\langle j|\hat\rho|j\rangle$. If we add to it that $\hat\rho$ is a pure state, i.e., $\hat\rho=|\psi\rangle\langle\psi|$, we get the further reduction to $P(A_j)=|\langle\psi|j\rangle|^2$, which everyone remembers from the textbooks.

To understand informational completeness, we examine the right-hand side of Eq.\ (\ref{BR}).  The trace of the product of two Hermitian operators $\hat{C}$ and $\hat{D}$ is actually an inner product $(\hat{C},\hat{D})$---not an inner product on the space ${\cal H}_d$ itself, but on the linear vector space of Hermitian operators ${\cal L}({\cal H}_d)$.  So, the Born Rule in Eq.~(\ref{BR}) boils down to simply taking the inner product of $\hat\rho$ and $\hat{A}_j$.  No fancy absolute-value-squared of an inner product, just an inner product itself.  The right-hand side of Eq.~(\ref{BR}) has a simple geometrical meaning:  It projects a vector $\hat\rho$ onto another vector $\hat{A}_j$.  Particularly, if we know the projections of a vector onto a complete basis for the vector space, then we can reconstruct the vector itself.

Putting these ideas together, if we can find any measuring device for which the $\{\hat A_j\}$ form a complete basis for the space ${\cal L}({\cal H}_d)$, then the probabilities $P(A_j)$ will uniquely specify a quantum state $\hat\rho$.  Of course, not all measuring devices fulfill this condition:  For instance a measurement of the old von Neumann variety cannot fulfill it because there are only $d$ elements to it, and the dimensionality of the space ${\cal L}({\cal H}_d)$ is $d^2$.  It thus takes $d^2$ vectors to span the space.  Nonetheless, such informationally complete measurements abound---they exist in all dimensions~\cite{Caves02b,DeBrota2021b}.  Once {\it any one such measurement}---say $\{\hat R_i\, |\, i=1,\ldots, d^2\}$---is singled out as being a {\it reference\/} measurement, one can rewrite all quantum states as the probability distribution it generates for its outcomes:
\be
P(R_i) = {\rm tr}\, \hat\rho\hat{R}_i \;.
\label{Pass}
\ee
With that correspondence, one can say that a quantum state {\it just is\/} a probability distribution!  This is naturally music to QBist\index{QBism} ears.

But at what cost?  We can have a new formalism for quantum theory, one that involves probabilities only, but does this transformation make the theory awkward or mathematically recalcitrant? For instance, what we have said so far establishes a one-to-one mapping (an injection) taking density operators and probability vectors,
\be
\hat\rho \quad \longrightarrow \quad \mathbf{P} = \Big[ P(R_1),\, P(R_2),\, \ldots,\, P(R_{d^2}) \Big]\;,
\label{Hermin}
\ee
so that we have a representation of quantum states as living in a probability simplex rather than a complex vector space, but we have not established that the mapping is onto (a surjection).  In fact, it is not, and regardless of the choice of our reference measurement it cannot be.  So instead of having the full probability simplex available for our $\mathbf{P}$, there will be a potentially very complicated convex subset of the simplex in which the $\mathbf{P}$'s reside.  Philosophically, that is OK, because we wanted to imagine the Born Rule as a means of giving us guidance in the first place.  Part of that guidance might be, ``Hey that $\mathbf{P}$ is a stupid choice; it doesn't live in the convex subset I was telling you about.''  Still, it does indicate that one might want to be very careful with the choice of the reference measurement so that, for instance, the allowed region for the $\mathbf{P}$ becomes as mathematically simple as possible, or so that the representation becomes as filled with guidance as possible, or perhaps both.

This brings us to the cutting edge of QBism and also sets the stage for giving the mathematically simplest form by which to take the Born Rule as a normative relation.  For all quantum systems ${\cal H}_d$ with $d=2, 3, \ldots, 196$ (and a scattering of other dimensions up to 45,372, so far), we know that there is a {\it very special\/} kind of informationally complete measurement $\{R_i\}$~\cite{Fuchs17,Scott17,Grassl24,Appleby2025,Bengtsson2025}---it goes by the name of {\it symmetric informationally complete\/} (or SIC) measurement.  \index{SIC-POVM} As the name implies, the operators $\hat R_i$ can be chosen to have a certain (amazing) symmetry.  Each can be written in terms of a pure quantum state $\hat\sigma_i=|\psi_i\rangle\langle\psi_i|$ via
\be
\hat R_i = \frac{1}{d} \hat\sigma_i \;,
\ee
while pairwise the $\hat\sigma_i$ obtain completely uniform inner products with each other:
\be
{\rm tr}\, \hat\sigma_i\hat\sigma_k = \frac{1}{d+1} \qquad \mbox{for all } i\ne k\;.
\label{RubyNell}
\ee
Why call this symmetry amazing?  Well, each $\hat\sigma_i$ is specified by $2d-2$ real parameters, and there are $d^2$ of them, so we are looking to find $\approx 2d^3$ parameters to specify the full set.  On the other hand, there are $\frac{1}{2}(d^4 - d^2)\approx \frac{1}{2}d^4$ equations of constraint.  Of course, these are nonlinear equations, but at first sight the system might appear far too over-constrained to have any solutions at all.  Expressing a sentiment of William Wootters, \index{Wootters, William K.} one might wonder whether such structures even have a ``right to exist.''

It is a long story to tell why the SICs \index{SIC-POVM} were first considered in proto-QBism~\cite{Caves99}.\footnote{Essentially, it was motivated by the struggle to prove a ``quantum de Finetti representation theorem''~\cite{Caves02b,Fuchs04} to make sense of quantum-state tomography, where the usually stated goal is to ``find the unknown quantum state'' \index{quantum de Finetti theorem} \index{quantum-state tomography} (an oxymoron in QBism). Ultimately the concept was not needed for the proof of the theorem, but it stayed around as a topic of interest and may end up being QBism's\index{QBism} greatest contribution to workaday physics. For instance, SIC states are now understood to have ``maximal magic'' \index{magic (quantum computation)} for quantum computation~\cite{Cuffaro2024}.} Since then, operator sets with this symmetry have become an object of study in their own right, with well more than 150 papers devoted to their further properties and structure~\cite{Fuchs17}. It is a very deep subject with mathematical tendrils running everywhere~\cite{Appleby11,Appleby15,Stacey2021}.  In the last few years, it has even become clear that the construction of these structures is related to one of the remaining unsolved problems proposed by David Hilbert \index{Hilbert, David} at the turn of the last century---Hilbert's 12th problem~\cite{Appleby2025}. \index{Hilbert's 12th problem} An unforeseen jewel buried within quantum theory! For our purposes, it is enough to know that most researchers in the quantum information community believe SICs exist in {\it all\/} finite dimensions. It just becomes a question of proving it (and reaping the spin-offs that might arise in the process).  {\it Meanwhile, let us suppose SICs do indeed exist in all finite dimensions.} We will show the pretty form the Born Rule \index{Born Rule} takes under this supposition.

Let us consider a {\it completely arbitrary\/} measurement $\{\hat E_j\}$. It might be of the old von Neumann variety or still something else entirely. If an agent had assigned a quantum state $\hat\rho$, then just as before, the Born Rule would specify
\be
Q(E_j)= {\rm tr}\, \hat\rho\hat{E}_j \;,
\label{Bijoux}
\ee
though now we use the notation $Q(E_j)$ for the probabilities. (The reason for the notational change will become clear shortly.)

Now, for a SIC measurement, \index{SIC-POVM} the operators $\hat R_i$ are used to calculate probabilities for its outcomes, but the projection operators $\hat\sigma_i$ from which the $\hat R_i$ are constructed can also be viewed as pure quantum states.  Thus, supposing a quantum state $\hat\sigma_i$ had been assigned to the quantum system for whatever reason, let us ask:  What would the Born Rule tell us of the probability for an outcome $E_j$?  It is ${\rm tr}\, \hat\sigma_i \hat E_j$ of course, as usual, but let us codify this with a name,
\be
P(E_j|R_i) = {\rm tr}\, \hat\sigma_i \hat E_j\;.
\label{Punt}
\ee
We do this because the quantity can be viewed as a conditional probability distribution for $E_j$ conditioned on $R_i$ if we consider the right scenario.  For instance, one might imagine the agent starting with the quantum state $\hat\rho$, performing a SIC measurement on the system, and then passing it on to the general measurement $\{\hat E_j\}$.  In light of the intermediate SIC measurement, the agent should update from $\hat\rho$ to some new quantum state, and if he is using L\"uders rule \cite{busch2009luders} to do this, it would be to $\hat\sigma_i$.  So, if the agent happens to know which click $R_i$ occurred in such a {\it cascaded measurement}, but not which $E_j$, he would indeed write down $P(E_j|R_i)$ as the conditional probability for the latter measurement outcome.

But what about the probability of the first click in this scenario?  It would be given Eq.~(\ref{Pass}), i.e., just the representation of $\hat\rho$ with respect to the SIC measurement in Eq.~(\ref{Hermin}).  Furthermore, because of the special symmetry in Eq.~(\ref{RubyNell}), there is an impressively clean reconstruction of $\hat\rho$ in terms of the $\hat\sigma_i$:
\be
\hat\rho = \sum_{i=1}^{d^2}\left[(d+1)P(R_i)-\frac{1}{d}\right]\! \hat\sigma_i\;.
\label{LillyLilac}
\ee
Now, recall that $\mathbf{P}$ cannot be just any probability in the probability simplex, but must be confined to a convex region of a certain variety.  For instance, one easy-to-see property of the region is that for all the $\mathbf{P}$ in it, $P(R_i)\le \frac{1}{d}$ for all the $i$.  So, for no quantum state $\hat\rho$, can one ever be too sure of the outcome of a SIC measurement.  In the end, the precise conditions on $\mathbf{P}$ so that it is consistent with all aspects of the Born Rule is simply that the $\hat\rho$ reconstructed in Eq.~(\ref{LillyLilac}) be positive semi-definite.  In other words, $\mathbf{P}$ must be in a convex set whose extreme points satisfy the condition of being pure quantum states. These extreme points can be expressed in two equations: one a quadratic, ${\rm tr}\, \hat\rho^2=1$, which in the language of $\mathbf{P}$ becomes simply an equation for a sphere,
\be
\sum_i p(R_i)^2 = \frac{2}{d(d+1)}\; ;
\label{Lucy}
\ee
and one a cubic, ${\rm tr}\, \hat\rho^3=1$, which becomes,
\be
\sum_{ijk} c_{ijk}\, P(R_i)P(R_j)P(R_k)=\frac{d+7}{(d+1)^3}\;,
\label{Desi}
\ee
where the real numbers $c_{ijk}={\rm Re} \big[{\rm tr}(\hat\sigma_i\hat\sigma_j\hat\sigma_k)\big]$ have impressive symmetry properties of their own~\cite{Appleby11,Appleby15,Stacey2021}.

We now have all the ingredients we need to write down the Born Rule \index{Born Rule} as expressed in Eq.\ (\ref{BR}), but purely in terms of probabilities.  Simply plugging the representation (\ref{LillyLilac}) into Eq.~(\ref{Bijoux}) and collecting terms gives~\cite{Fuchs13a}:
\be
Q(E_j) = \sum_{i=1}^{d^2}\left[(d+1)P(R_i)-\frac{1}{d}\right]\! P(E_j|R_i)\;.
\label{TheOneRing}
\ee
And that's what we've been after.\footnote{Of course, this form supposes that a SIC always exists for any dimension $d$.  But even if SICs don't always exist, we could still give a QBist\index{QBism} explanation of the Born Rule by using one of those abundant (unsymmetrical) informationally complete POVMs mentioned above~\cite{DeBrota2021b,DeBrota2020b}, which {\it are\/} known to always exist.  It is just that such a version of the Born Rule will not be so simple as this one.}

Notice that, perhaps contrary to expectation,
\be
Q(E_j) \ne P(E_j) \equiv\sum_{i=1}^{d^2}P(R_i) P(E_j|R_i)\;,
\label{Loser}
\ee
where the right-hand side is simply an expression of the Law of Total Probability. This discrepancy, or ones morally equivalent to it (as for instance Feynman's \index{Feynman, Richard P.} discussion of the double slit experiment), have often been advertised as a statement that probability theory itself must be modified when the concern is quantum phenomena~\cite{Feynman51}.  But QBism sees Eq.~(\ref{TheOneRing}) as strictly an {\it addition\/} to probability theory---nothing about standard probability theory is negated.  The reason there is no equality in Eq.~(\ref{Loser}) is because the pure $\{\hat E_j\}$-measurement and the cascaded measurement (i.e., first the $\{\hat R_i\}$ then the $\{\hat E_j\}$) refer to two different hypothetical scenarios, ones that cannot be performed simultaneously.  In that sense, they echo Bohr's \index{Bohr, Niels} \index{complementarity} own usage of the term ``complementary''~\cite[p.\ 57]{Bohr49-1}:
\bq\small
\noindent In fact, we must realize that in the problem in question we are not dealing with a {\it single\/} specified experimental arrangement, but are referring to {\it two\/} different, mutually exclusive arrangements.
\eq

One way to think of the reference SIC measurement is that it stands above all other quantum measurements---it is a single measurement that is complementary to all the others.  No matter what the $\{\hat E_j\}$, the $\{\hat E_j\}$ and $\{\hat R_i\}$ measurements cannot be made together.  This is ultimately the reason for Eq.~(\ref{TheOneRing}).  One might say that if a value for $\{\hat E_j\}$ is allowed to come into being, a value for $\{\hat R_i\}$ cannot be assumed sitting there unrevealed.  If it were otherwise, we would have equality in Eq.~(\ref{Loser}):  The $R_i$ would then be an ontic-variable model for the $\{\hat E_j\}$ in the sense of Harrigan and Spekkens~\cite{Harrigan10}, and we would have used the standard Bayesian Law of Total Probability \index{Law of Total Probability} to calculate $Q(E_j)$.  Instead we only use the ingredients from that law, but in the new combination that is uniquely quantum---Eq.~(\ref{TheOneRing}).

\begin{figure}
\begin{center}
    \includegraphics[width=4.5in]{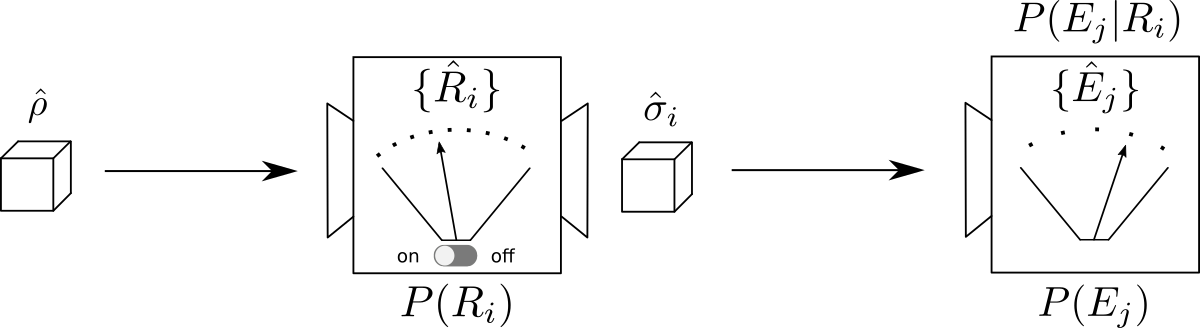}
\end{center}
\begin{center}
    \includegraphics[width=4.5in]{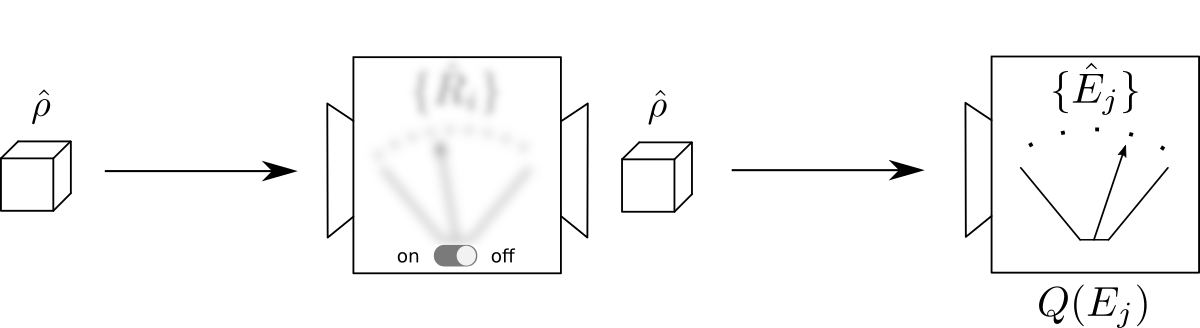}
\end{center}
\caption{\small Two distinct experiments.  In QBism,\index{QBism} the Born Rule \index{Born Rule} is not about either one individually, but rather about the connections between their probabilities.  In the top experiment, the reference device is turned on so that there are three probabilities in its telling: $P(R_i)$, $P(E_j|R_i)$, and $P(E_j)$. They must satisfy the Law of Total Probability, \index{Law of Total Probability} i.e., the right-hand side of Eq.~(\ref{Loser}). However, in the bottom experiment the reference device is turned off---there is only one probability $Q(E_j)$ in its story. The Born Rule is the narrative glue that ties the two stories together.}
\label{Metzenbaum}
\end{figure}

As a normative relation, what the Born Rule Eq.~(\ref{TheOneRing}) is doing, is suggesting to the agent:  Before you gamble on the outcomes of this experiment, you should think hard about how you would gamble on the cascaded one as well. If you come up with sets of probabilities such that
\be
Q(E_j) \ne \sum_{i=1}^{d^2}\left[(d+1)P(R_i)-\frac{1}{d}\right]\! P(E_j|R_i)\;,
\ee
then you should strive harder to find a set of probability assignments which instate equality instead.  You should dip back into your experience and reassess: This is the only suggestion of the Born Rule.  It doesn't single out any of the terms as more important than the others:  It might be $\mathbf{P}$ that the agent adjusts, or $\mathbf{Q}$, or some of the conditional probabilities, or some combination of all the above.  It is the relation that is normative, not the terms within it. As L. J. Savage \index{Savage, Leonard J.} says of probability theory~\cite{Savage54}:
\bq\small
\noindent
According to the personalistic view, the role of the mathematical theory of probability is to enable the person using it to detect inconsistencies in his own real or envisaged behavior. [\ldots\!] An inconsistency is typically removable in many different ways, among which the theory gives no guidance for choosing.
\eq
QBism\index{QBism} says likewise of the Born Rule.  Though here the inconsistency is not of a purely logical kind, but between an agent's desire to flourish in the world in which he is immersed and that world's actual character.  

\subsubsection{QBism is No Copenhagen Interpretation, Nor Is It Many-Worlds}
\label{Haribo}

Many who encounter QBism for the first time are surprised by its reading of the Born Rule.  As we have emphasized, the usual reading is that a quantum state $\hat\rho$ is \emph{given} and a specification of a measurement $\{\hat E_j\}$ is \emph{given}, and only thereafter one calculates a probability distribution $Q(E_j)$.  But in Eq.~(\ref{TheOneRing}), all the terms have co-equal status:  They are all personalist Bayesian probability assignments.  Nonetheless the correspondence of the terms is clear: The collection $\{P(R_i)\}$ uniquely specifies a state $\hat\rho$ and the collection $\{P(E_j|R_i)\}$ uniquely specifies a measurement $\{\hat E_j\}$.  The former is old hat and well-known about QBism by now.  However the implications of the latter still seem not so widely recognized despite being made in print as early as 2002~\cite{Fuchs02a}.

\begin{figure}
\begin{center}
\includegraphics[width=3.5in]{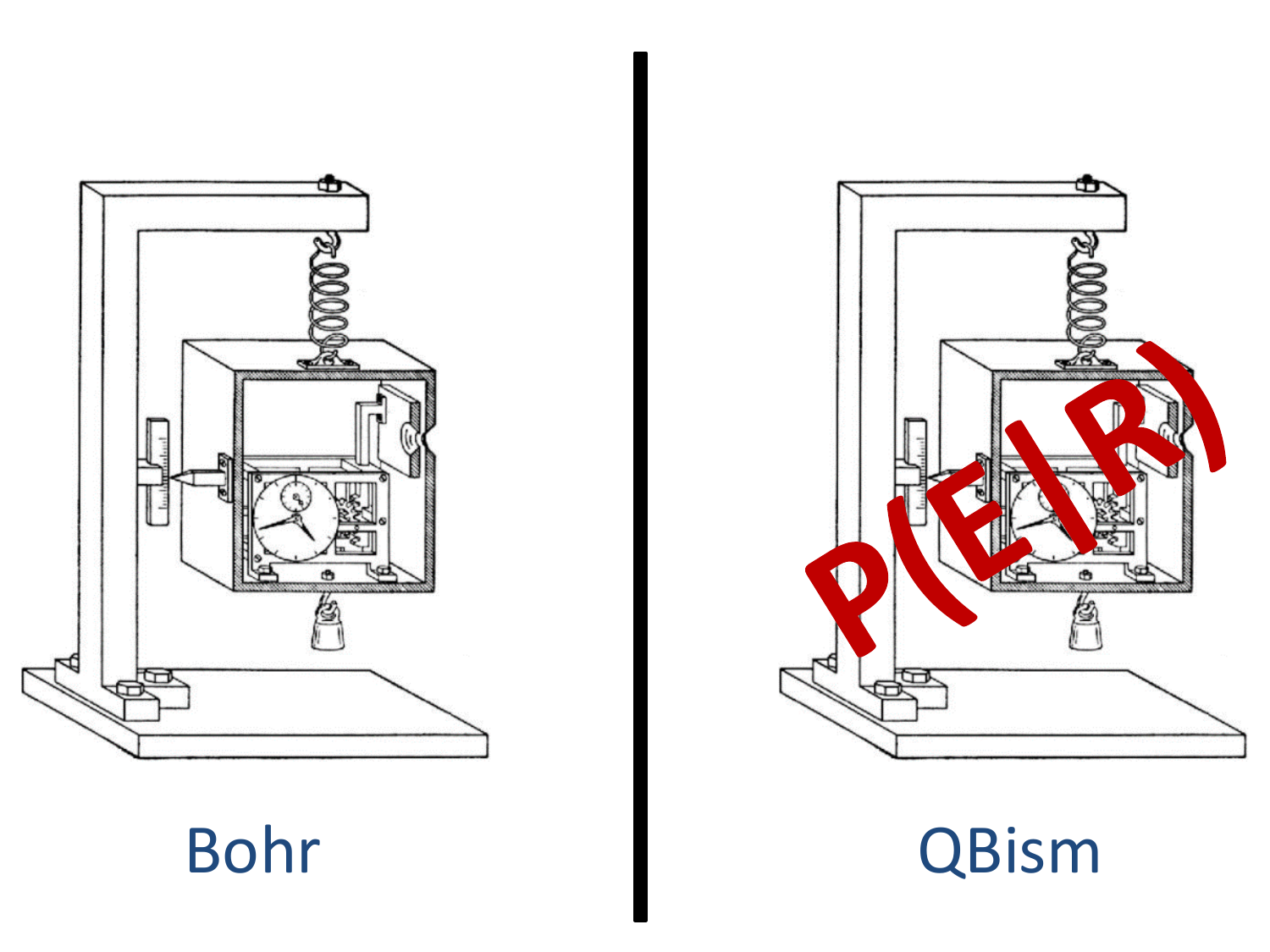}
\end{center}
    \caption[Caption for LOF]{\label{BohrAndQBism} \small Bohr \index{Bohr, Niels} would often depict quantum measuring devices as heavy, bulky instruments, firmly bolted to their laboratory benches. But the QBist realizes that using these measuring devices means implicit and explicit statistical analyses deeply dependent upon an agent's priors. The residue of that analysis ultimately boils down to a specification of the $P(E_j|R_i)$ in Eq.~(\ref{Punt}). Who could walk into the laboratory and see the personal probabilities?}
\label{Scissors}
\end{figure}

Particularly, the personalist character of the $\{P(E_j|R_i)\}$ puts an absolute wedge between QBism\index{QBism} and Niels Bohr's \index{Bohr, Niels} way of thinking of quantum measurement~\cite{Fuchs2017}.  Consider Fig.~\ref{Scissors}.  As it illustrates, Bohr often depicted measuring devices as heavy, classical things firmly bolted to their laboratory benches. Surely the reason he did was because, as he emphasized over and over~\cite[p.\ 39]{Bohr49-1},
\bq\small
\noindent [H]owever far the phenomena transcend the scope of classical physical explanation, the account of all evidence must be expressed in classical terms.  The argument is simply that by the word ``experiment'' we refer to a situation where we can tell others what we have done and what we have learned and that, therefore, the account of the experimental arrangement and of the results of the observations must be expressed in unambiguous language with suitable application of the terminology of classical physics.
\eq
In effect, Bohr required that science be done by sharing blueprints.

By contrast, the best a QBist\index{QBism} experimenter can do is reveal to her colleagues her personal probability assignments for her personal experiences consequent upon her personal actions on a system. {\it What is another experimenter to do with that!?}  QBist Alice is never communicating a blueprint to QBist Bob!  She is neither offering a fact nor a proposition with a truth value in describing an experiment.  Instead, Alice offers a representation of her personal beliefs; Bob combines it in one way or another with his previously (partially) coherent belief system to make something new~\cite{Lindley82} and then offers the resultant back. So it goes, ping-pong, ping-pong, with each participant's query of the other a quantum measurement in its own right. Amanda Gefter~\cite{Gefter24} puts the central issue nicely in her criticism of Bohr's ``unambiguous language'' \index{language, unambiguous} policy. \index{Gefter, Amanda} Quoting Martin Buber, ``[I]t is not the unambiguity of a word but its ambiguity that constitutes living language,'' she adds,
\bq\small
\noindent [Agreements] are achieved through participatory sense-making, not by compelling two agents to agree on some pre-existing facts in the world, but by allowing two
agents to forge new ``f/acts'' through \ldots\ interactions that are always provisional, always renegotiable, always vulnerable to breakdown, and as such genuine and real.
Agreement is never guaranteed, but it can be enacted. \index{intersubjective agreement}
\eq
In contrast to Bohr, in QBism,\index{QBism} it's ambiguity all the way down.

Finally, another insight can be gleaned from our formalism to do with a completely different interpretative direction than Bohr's:  Everett's \index{Everett, Hugh III} many-world theory. For, just as one can see that the specification of a measuring device is as ``diaphanous'' as any personalist Bayesian probability assignment, one can see the same thing for unitary evolution operators---the very fuel the Everett interpretation runs on.\footnote{We speak of ``the'' Everett interpretation for simplicity, even though we have never met two Everettians who fully agree on how to make it work.}

Let us start with an agent who assigns a quantum state $\hat\rho$ to a system at time $t_0$ and a time-evolved state $\hat U\hat\rho\,\hat U^\dagger$ (for some unitary $\hat U$) at time $t_1$.  How would we put this in normative terms?  We would do as we have before:  Introduce a SIC reference device and reexpress the initial quantum state as
\be
P_{t_0}(R_i) = \tr \hat\rho\hat{R}_i
\ee
and another other probability distribution at time $t_1$,
\be
P_{t_1}(R_j) = \tr \big(\hat U\hat\rho\,\hat U^\dagger\big)\hat{R}_j = \tr \hat\rho\,\big(\hat U^\dagger\hat{R}_j \hat U\big)\;.
\ee
This suggests that we focus on the measurement $\hat{R}^\prime_j = U^\dagger\hat{R}_j\hat U$ and think of it just as any of the measurements we could put into the right side of Fig.\ \ref{Metzenbaum} to get:
\be
P_{t_1}(R_j)=Q(R^\prime_j)\;.
\ee
With this it becomes obvious what the relation between the $P_{t_0}(R_i)$ and $P_{t_1}(R_j)$ must be.  It is just the Born rule as expressed in the language of Eq.\ \eqref{TheOneRing}, but with slightly modified variables:
\be
P_{t_1}(R_j)=\sum_{i=1}^{d^2}\left[(d+1)P_{t_0}(R_i)-\frac{1}{d}\right]\! P(R^\prime_j|R_i)\;,
\label{timetrip}
\ee
where
\be
\label{RiffRaff}
P(R_j^\prime|R_i)=\tr \hat\sigma_i \hat{R}_j^\prime\;,
\ee
With this, {\it the very meaning of unitary evolution\/} is at hand: It is captured all and only by the conditional probability assignments in Eq.\ \eqref{timetrip}.  That is, it is transparently as much a personal judgment as the quantum states assignments $P_{t_0}(R_i)$ and $P_{t_1}(R_j)$ are.  

What this means for any attempt to shoehorn QBism\index{QBism} into an Everettian worldview---as some Everettian colleagues have expressed a desire to do\footnote{As an example, Sean Carroll \index{Carroll, Sean M.} said in a recent podcast~\cite{CarrollPodcast}, ``[M]y current conjecture \ldots\ is that QBism is an effective coping strategy for people who don't want to accept many worlds, and they'll eventually get there. The Bayesianess of it is absolutely in common.'' Such a superficial understanding of QBism!} simply because they {\it say\/} their probabilities are Bayesian too---cannot be done.  To get off the ground, the Everett interpretation requires an {\it ontic\/} universal unitary evolution and a means through it to define decoherence, branching, and the conditions under which probabilities can even be assigned. There is no room for the \index{van Fraassen, Bas C.} van-Fraassen-style \index{will to believe} ``will to believe'' considerations discussed at the end of Section~\ref{FirstTenet}.  More poignantly, there is the simple fact that in QBism, decoherence \index{decoherence} goes the other way around:  It is a kind of subjective judgment an agent can make~\cite{Fuchs2012,DeBrota2023}, nothing ontic about it.

\subsection{Second Tenet: My Probabilities Cannot Tell Nature What To Do}
\label{TenetTwo}

Suppose someone believes with all their heart that their partner will never cheat on them. In relationships, sometimes beliefs are that strong. Yet no matter how strong the belief, we know that the partner need not comply---tragedy sometimes happens.  The person's beliefs or gambling commitments are one thing; the events of the world are another.  This is what QBists\index{QBism} mean by saying that a quantum state has no ``ontic hold'' on the world.  An agent's mesh of beliefs may cause them to make a pure quantum-state assignment for some quantum system, and they could even consider a yes-no measurement that is exactly the projection onto that state.  Using the Born Rule, \index{Born Rule} the agent would then calculate the probability of a ``yes'' outcome to be {\it exactly\/} 1---i.e., they should believe a ``yes'' will occur with all their heart.  But that does not mean that the world must comply.  The agent's quantum state assignment does not mean that the world is {\it forbidden\/} to give them a ``no'' outcome for this measurement~\cite{Caves07}.  All {\it they\/} know by using the Born Rule is that {\it they\/} have made the best gamble {\it they\/} could in light of all the other gambling commitments {\it they\/} might be making with regard to other phenomena (other experiments, etc.).

There is a sense in which this unhinging of the Born Rule \index{Born Rule} from being a ``law of nature'' in the usual conception~\cite[pp.~86--101]{Weinberg15}---i.e., treating it as a normative statement, rather than a descriptive one---makes the QBist notion of quantum {\it indeterminism\/} a far more radical variety than anything proposed in the quantum debate before.  It says in the end that nature does what it wants, without a mechanism underneath, and without any ``hidden hand'' \cite{Fine89} of the likes of \index{von Mises, Richard} von Mises's {\it Kollective\/} or \index{Popper, Karl} Popper's {\it propensities\/} or \index{Lewis, David} Lewis's {\it objective chances\/}, or indeed any conception that would diminish the autonomy of nature's events.  Nature and its parts do what they want, and we as free-willed agents do what we can get away with.  Quantum theory, on this account, is our best means yet for hitching a ride with the universe's pervasive creativity and doing what we can to contribute to it.

\subsection{Third Tenet: A Measuring Device Is Literally an Extension of the Agent}

In the discussion of the Tenet 2, we deliberately made it hard to not notice how often the word {\it they\/} was emphasized.  This was given a third-person perspective for presentational purposes, but the message was intended to be about the first-person.  This is because QBism's understanding of quantum theory is purely in first-person terms:  ``When I---the agent---write down a quantum state, it is {\it my\/} quantum state, no one else's.  When I contemplate a measurement, I contemplate its results, outcomes, consequences {\it for me}, no one else---it is {\it my\/} experience.''  This is why it was declared that any measuring equipment should literally be considered part of the agent.  But of course, it is the same for {\it you\/} when {\it you\/} are the one applying quantum theory.  It is what it means for quantum theory to be a ``user's manual'' or handbook for each of us~\cite{Varenna17}.

The roots of this idea come from Heisenberg~\cite{Heisenberg35} and Pauli, but go much further. \index{Heisenberg, Werner} \index{Pauli, Wolfgang} \index{Bohr, Niels}  In a 1955 letter to Bohr, Pauli wrote~\cite{Pauli55},
\bq\small
\noindent
As it is allowed to consider the instruments of observation as a kind of prolongation of the sense organs of the observer, I consider the impredictable change of the state by a single observation \ldots\ to be {\it an abandonment of the idea of the isolation (detachment) of the observer from the course of physical events outside himself}.
\eq
So the idea was certainly in the air with some of the Copenhageners, but one has to wonder how seriously they took themselves.  By 1958 Pauli seemed to be on the retreat~\cite{Pauli58a}:
\bq\small
\noindent [P]ersonal qualities of the observer
do not come into the theory in any way---the observation can be made by
objective registering apparatus, the results of which are objectively
available for anyone's inspection.
\eq

QBism,\index{QBism} in contrast, takes the idea of ``the instruments of observation as a \ldots\ prolongation of the sense organs of the observer'' deadly seriously and runs it to its logical conclusion.  This is why QBists opt to say that the {\it outcome\/} of a quantum measurement is a {\it personal experience\/} for the agent gambling upon it.  Whereas Bohr always had his classically describable measuring devices mediating between the registration of a measurement's outcome and the individual agent's experience, for QBism the outcome {\it just is\/} the experience.  This is the aspect of QBism that takes it firmly into the territory of early American pragmatism~\cite[Ch.\ 7]{Jay2005}.

The reason for this move has to do with the desire to maintain quantum theory as a universally valid schema:  a handbook anyone can use, in principle, for better gambling on the consequences of any action on {\it any\/} physical system.  This forces a reconsideration of the Wigner's friend paradox, \index{Wigner, Eugene P.} \index{Wigner's friend} a place many Copenhageners don't seem to want to go~\cite{Englert13}.  The key realization is that going to this length is required for the consistency of QBism:  When Wigner writes down a quantum state for {\it any\/} quantum system---say, the compound system of an electron plus his friend, who himself is believed to be performing a spin-measurement on the electron---Wigner is {\it never\/} telling a story~\cite[p.\ 592]{Timpson08} of what is happening on the inside of the conceptual box surrounding these things~\cite[Section III]{Fuchs10a}\@.  Nor is the unitary evolution he ascribes to the conceptual box telling a story of the actual goings-on within it---a point we argued for in Section~\ref{Haribo}.  

Wigner's quantum-state assignment and unitary evolution for the compound system are only about his {\it own\/} expectations for his {\it own\/} experiences should he take one or another action upon the system or any part of it.  One such action might be his sounding the verbal question, ``Hey friend, what did you see?,'' which will lead to one of two possible experiences for him.  Another such action could be to put the whole conceptual box into some amazing quantum interference experiment, which would lead to one of two completely different experiences for him.  The friend on the inside, of course, has his own story to tell, but that's his business, not Wigner's.  In fact, the closest QBism\index{QBism} comes to telling any story at all is to say, whatever is in the conceptual box---whatever is happening in there---Wigner would be well-advised to adopt the Born Rule \index{Born Rule} for gambling upon its potential impact to {\it him}~\cite{DeBrota2020a}.

When Bohr \index{Bohr, Niels} writes of ``the factual character of the observations'' that ``can be communicated to everyone''~\cite{Folse85,Faye91}, QBism asks, {\it ``But then what about Wigner's friend?,''} and so it goes into the usual debates.  Still perhaps there is one issue to consider that breaks some new ground.  In Wigner's \index{Wigner, Eugene P.} original argument, he wrote ``If the atom is replaced by a conscious being, the wave function $|\Phi\rangle$ \ldots\ appears absurd because it implies that my friend was in a state of suspended animation before he answered my question.''  Suspended animation?  Is it a necessary consequence of QBism's\index{QBism} response to Bohr that a pure quantum-state assignment leads to a kind of zombie-hood for the friend?

The difficulty comes from trying to understand what a pure-state assignment entails for an agent who makes it for an external physical system, which in another context, might be thought of as another agent making her own quantum-state assignment about the first. In all other quantum interpretations, there is always a move (even if only implicit) to treat one agent as a ``super-observer''~\cite{Brukner17} and the other as a ``zombie.''  But QBism can have none of that, else its whole foundation would crumble: QBism's very selling point is that it takes quantum theory as an addition to decision theory, and consequently all agents must be on an equal footing.  This is where a deep confusion could arise. It comes from the lingering habit of thinking that a pure-state assignment for a system means that one has a God-like control of the system itself, that one can set its inner workings to be whatever one wants.

From a QBist\index{QBism} perspective, there is no contradiction in calling a physical system ``a person''  (or an agent) at the same time as writing down a pure quantum state for it.  Take some system that a QBist agent believes will pass a \index{Turing test} Turing test, Danny Greenberger's \index{Greenberger, Daniel M.} ``big red button'' test~\cite{Greenberger2014}, one for which she feels a heartfelt empathy, one for which she would feel the loss of their companionship should they leave, etc., etc.  One can consider a very long list of such things.  These are all beliefs that must be rolled into a QBist's quantum state assignment for the system:  They are not thoughts to be ignored, but rather ones to be refined until they fit within a numerical framework capturing the agent's best gambles.  Perhaps it so transpires that this process of refinement leads to some pure-state assignment. Well then, so what? It only means that the pure-state assignment landed on will contain all those beliefs. The agent's original belief that the other person is her friend is incorporated into the state assignment, not overridden by it.

\section{Bell-type Inequalities and Locality}

\begin{flushright}
\baselineskip=13pt
\parbox{4.0in}{\baselineskip=13pt\footnotesize
The kinematically independent parts into which a system can be resolved need not be spatially separated, nor need they even refer to different particles.
}\\
\footnotesize --- Hermann Weyl \cite{Weyl1950} \index{Weyl, Hermann} \\
\end{flushright}

We have spoken of an ontological lesson arising from the three tenets of QBism.\index{QBism} Could it be as a consequence of our way of thinking that Tim Maudlin's \index{Maudlin, Tim} claim in Ref.~\cite{Maudlin14}---``What Bell \index{Bell, John S.} proved, and what theoretical physics has not yet properly absorbed, is that the physical world itself is non-local.''---is actually true?  It is just the opposite, and it is worthwhile seeing how his kind of thinking clashes with QBism at more than one level.  

First, as Maudlin forewarns: ``Some people, hell-bent on denying the conclusion of the EPR argument, have taken to thinking they just have to reject this EPR criterion. If this can be done, then the conclusion [of non-locality] need not be accepted.''  We wouldn't say that QBism was ever ``hell-bent on denying EPR''---it likely wasn't even on QBism's radar screen in the early days (CAF does not remember it to be so)---but the denial of the EPR reality criterion \index{EPR criterion of reality} is in fact one of QBism's conclusions~\cite{Fuchs10a}. This is because by Tenet 2 even probability-1 assignments are nothing more than personal gambling commitments. QBism thus naturally spurns the EPR criterion, and this blocks Maudlin's too-hasty inference from the violation of Bell-type inequalities \index{Bell inequality} to ``non-locality''~\cite{Fuchs14,Varenna17}. 

More interesting, however, is to see how the allure of Bell's thinking can be dispelled through a simple technical consideration to do with Tenet 1\@.  For this, we deploy it on a {\it single\/} quantum system---one right in front of the agent, with no spatially separated parts whatsoever.  As an example, focus on a four-level quantum system, a so-called ``ququart''~\cite{Seifert2023}. This could be a harmonic oscillator restricted to its first four energy levels, or for instance a single spin-3/2 particle. All systems of this kind have an associated 4-dimensional \index{Hilbert space} Hilbert space, ${\cal H}_4$.

This Hilbert space is isomorphic to the tensor product of two qubit-sized Hilbert spaces: ${\cal H}_4 \simeq {\cal H}_2 \otimes {\cal H}_2$. Why not treat it as such? This can always be done---technically by separating its algebra of operators into two commuting subalgebras, \index{Tsirelson's theorem} a result known as Tsirelson's theorem~\cite{Tsirelson1993}.  There are in fact an infinite number of ways of doing this, but let us focus on one such decomposition.  What is important is that with respect to this decomposition of the algebra, there will be every kind of phenomena one usually associates with bipartite spatially-separated systems.  There will be unentangled states and maximally entangled states (as usually discussed in Bell scenarios) and everything in between.  This is a pedestrian consequence of how the tensor product of vector spaces is defined. Moreover, any measurements on the ``component'' qubits separately will correspond to commuting measurements on the full ${\cal H}_4$ system---so they could be performed in any order or even simultaneously, and the statistics of their outcomes will remain the same.

Suppose now that Alice has a large supply of the ${\cal H}_4$ systems, all of which she has ascribed the same, maximally entangled state.  Then on each, she can imagine performing a two-step measurement:  First she measures an operator on one qubit, then an operator on the other, where the choices of measurements she makes are determined by the desire to see a Bell inequality violation. For example, they could be the operator pairs $\hat A^{(1)}$, $\hat A^{(2)}$ and $\hat B^{(1)}$, $\hat B^{(2)}$ that appear in the Bell--CHSH inequality. Alice will then predict with high probability that she will indeed find a set of correlations $Q\big(A^{(1)}_k,B^{(1)}_l\big)$, $Q\big(A^{(1)}_m,B^{(2)}_n\big)$, etc., that give exactly such a violation.

Should she be surprised by this result?  Hardly! For she knows that it is just a consequence of using the Born Rule, as for instance represented in Eq.~(\ref{TheOneRing}).  If she were a QBist,\index{QBism} she would never question the \emph{strict adherence} to the Born Rule \index{Born Rule} for every quantum calculation.  Here the culprit for a Bell inequality violation is obvious:  It arose from nothing more than a {\it silly\/} and {\it completely unjustified\/} comparison between the Born Rule and a collection of {\it ad hoc\/} rules of the form
\bea
\label{DoofusRule}
Q\big(A^{(1)}_k,B^{(1)}_l\big) &=& \int P\big(A^{(1)}_k|\lambda\big)\,P\big(B^{(1)}_l|\lambda\big)\,P(\lambda)\, d\lambda
\\
Q\big(A^{(1)}_m,B^{(2)}_n\big) &=& \int P\big(A^{(1)}_m|\lambda\big)\,P\big(B^{(2)}_n|\lambda\big)\,P(\lambda)\, d\lambda
\eea
and so on for the other combinations of qubit observables---namely, a ``local'' hidden variable model with respect to the tensor decomposition.  A QBist\index{QBism} would never do that for the case of a ququart, and no one of any other interpretational bent would either.  It would be, as we said already, silly.

Now, consider the usual Bell scenario, where Alice and Bob are spatially separated.  In this case, those other interpretational bents find the disparity between Eqs.~(\ref{TheOneRing}) and (\ref{DoofusRule}) remarkable and in need of further physical explanation.  A QBist\index{QBism} would say there is nothing deeper here than in the first story. The algebra of a four-level system is the same as that for a pair of qubits. In the latter case, people manage to fool themselves into thinking that ``locality'' requires a joint probability assignment of the same {\it ad hoc\/} type dismissed above. But ``locality'' makes no such demands. By Tenet 3, what an agent \emph{causes} by direct participation happens locally; by Tenet 2, any state change induced at a distance is not an ontic change. Due to the no-signaling principle, which follows ultimately from the Born Rule \index{Born Rule} itself, no state change made in reaction to a measurement requires the introduction of ontological elements that spirit information away superluminally~\cite{Stacey2021b}. Why all the fuss over ``non-locality'' then? Perhaps there is a psychological mystery, but there is not a physical one. The genuine mystery, the one that should occupy our thoughts, QBism claims, is \emph{why the Born Rule in the first place?}

\section{Toward the Future}
\label{Future}

\medskip

\begin{flushright}
\baselineskip=13pt
\parbox{4.0in}{\baselineskip=13pt\footnotesize
It is difficult to escape asking a challenging question. Is the
entirety of existence, rather than being built on particles or fields
of force or multidimensional geometry, built upon billions upon
billions of elementary quantum phenomena, those elementary acts of
``observer-participancy,'' \index{observer-participancy} those most ethereal of all the entities
that have been forced upon us by the progress of science?
}\\
\footnotesize --- John Archibald Wheeler \cite{Wheeler82c} \index{Wheeler, John Archibald} \\
\end{flushright}

What does QBism say existence is built upon? QBism\index{QBism} should be seen as a project, not a final answer---so anything said here can only be tentative.  With that as a caution, let us give it a shot. The \emph{formalism} of quantum mechanics presupposes an agent, for quantum mechanics is a decision theory according to QBism.  But we would not call that an ``ontology of the agent.''  So far as we know, the raw world may have never contingently evolved any agents, much less users of quantum mechanics.  But granted that we do use quantum mechanics, we think there is an ontological lesson to be learned from it.

Speculatively, what QBism is aiming to make precise is that the ultimate ``stuff of the world'' is analogized to what we see in the act of quantum measurement: A temporary suspension of the subject-object distinction which gives rise to something new and \emph{sui generis} in the world~\cite{Fuchs2022}. Metaphorically, think of a quantum measurement as being like a marriage with the birth of a child as a result.  The key here, though, is the important word ``analogized.'' For what one really wants is a de-anthropocentrized notion of the essence of quantum measurement, so that one can see that it is going on ``everywhere, all the time''~\cite{Bell1990} (with and without \emph{Homo sapiens}).  To riff on Wheeler's suggestion, ``Might the universe be \emph{all} big bang?  Or so to speak, big bang all the way down?''  Might the stuff of creation be ``little bangs'', elementary acts of inter-participancy, \index{inter-participancy} of which \emph{observer}-participancy \index{observer-participancy} is only a special case?

What does QBism have to go on for building such a vision beyond the airy hints of Wheeler?  Actually, there is more than one philosophical tradition which QBism might build upon---from American pragmatism~\cite{Fuchs11a,MyStruggles} to certain strains of phenomenology~\cite{Berghofer2024,Bitbol2020}, \index{pragmatism} \index{phenomenology} with perhaps a stop at Whitehead \index{Whitehead, Alfred North} in between~\cite{Weber2006}.  Take for instance, \index{James, William} William James's premonition of both a ``big bang'' and the ``little bangs'' in {\sl Some Problems of Philosophy}~\cite{James40}:
\bq\small
\noindent
It is a common belief that all particular beings have one origin and source, either in God, or in atoms all equally old.  There is no real novelty, it is believed, in the universe, the new things that appear having either been eternally prefigured in the absolute, or being results of the same {\it primordia rerum}, atoms, or monads, getting into new mixtures.  But the question of being is so obscure anyhow, that whether realities have burst into existence all at once, by a single `bang,' as it were; or whether they came piecemeal, and have different ages (so that real novelties may be leaking into our universe all the time), may here be left an open question \ldots
\eq
Beyond all else, James stood for the idea that the universe is ``on the make''---a world that is ``ever not quite''~\cite{Stacey2023b}. What he sought was a vision of existence where {\it novelty\/} itself is the crucial ingredient.  How well does this fit with the kind of extreme indeterminism uncovered by QBism's Second Tenet?  This is an example of the kind of work left for our future, but the program is broader still.

There is a rumour that QBism is only about subjective probabilities and agents---nothing to do with saying something about what reality is.  But we hope we have already said enough to dispel that.  What is usually not appreciated is that the technical program of QBism is about re-expressing the quantum formalism in a way that makes manifest its purpose as providing aid to decision making agents immersed (and participating) in a world of this character---a world in constant creation~\cite{Fuchs2007}.  We believe that figuring out why the quantum formalism is tuned exactly the way it is, is the first step toward saying something precise about this kind of reality.

A clean example of this is the reformulation of the Born Rule \index{Born Rule} given in Eq.~(\ref{TheOneRing}).  QBists view that expression as not the end of a quest, but the starting point of a new one.  It is our surest hint yet for how to quantify the minimum ``participatoriality'' \index{participatoriality} entailed for agents immersed in a quantum world---indeed, in Ref.~\cite{Fuchs2023} its form is argued to be QBism's analog to Bohr's \index{Bohr, Niels} considerations about the role of $\hbar$ \index{quantum of action} \index{Planck's constant} in quantum measurement.  So, where better to start the quest to quantify a de-anthropocentrized notion of \index{inter-participancy} inter-participancy---the ontological residue QBism seeks?

This is not the place to explain the technical program of infrastructure-building QBism has undertaken for those next steps~\cite{Fuchs13a,Fuchs11b,Appleby09a,Appleby16b,Weiss2023,DeBrota2021,Stacey2023}, but it seems appropriate to end a contribution to a volume on ``100 Years of Ongoing Quantum Interpretation'' by saying a little bit about how QBism sees the next 100 years.  It's about squeezing a worldview from our current best theory of physics.  With some chance, it will match up with William James's \index{James, William} grand vision so beautifully expressed by \index{Durant, Will} Will Durant~\cite[p.\ 673]{Durant06}:
\bq\small
\noindent The value of a [Jamesian pluriverse], as compared with a universe, lies in this, that where there are cross-currents and warring forces our own strength and will may count and help decide the issue; it is a world where nothing is irrevocably settled, and all action matters. A monistic world is for us a dead world; in such a universe we carry out, willy-nilly, the parts assigned to us by an omnipotent deity or a primeval nebula; and not all our tears can wipe out one word of the eternal script. In a finished universe individuality is a delusion; ``in reality,'' the monist assures us, we are all bits of one mosaic substance. But in an unfinished world we can write some lines of the parts we play, and our choices mould in some measure the future in which we have to live. In such a world we can be free; it is a world of chance, and not of fate; everything is ``not quite''; and what we are or do may alter everything. 
\eq
How this gives hope to each of our roles in the world~\cite{Fuchs11a,Fuchs02b}. James and Durant provide the eloquence. The QBism of the future hopes to provide the precision.

\bigskip\bigskip
\noindent {\Large \bf Acknowledgements}\bigskip

\noindent This research was supported by the National Science Foundation through grants NSF-2210495 and OSI-2328774.








\end{document}